\documentclass[12pt]{iopart}
\usepackage{graphicx} 
\usepackage{cite}
\usepackage{xcolor}

\begin{document}

\title[Watt-level ultrafast 1.75 {\textmu}m laser system based on thulium-doped core and terbium-doped cladding \dots]{Watt-level ultrafast 1.75 {\textmu}m laser system based on thulium-doped core and terbium-doped cladding fluoride fibers}


\author{Dina Grace Banguilan$^1$, Yuto Iye$^1$, Kazuhiko Ogawa$^2$,\\ Eiji Kajikawa$^2$ and Takao Fuji$^1$}
\address{$^1$Laser Science Laboratory, Toyota Technological Institute, 2-12-1 Hisakata, Tempaku-ku, Nagoya, 468-8511, Japan}
\address{$^2$FiberLabs Inc., KDDI Laboratories Building, 2-1-15 Ohara, Fujimino, Saitama 356-8502, Japan}
\ead{dcbanguilan@toyota-ti.ac.jp}

\vspace{10pt}
\begin{indented}
\item[] February 16, 2026
\end{indented}

\begin{abstract}

We report watt-level femtosecond pulses in the 1.75~{\textmu}m region using a thulium-doped core, terbium-doped cladding fluoride (Tm:Tb:ZBLAN) fiber laser
system. 
The seed pulse is generated through stimulated Raman scattering in a silica fiber pumped by a femtosecond erbium-doped fiber laser. 
The soliton is stretched with a chirped fiber Bragg grating (CFBG), and is subsequently amplified through a multi-stage Tm:Tb:ZBLAN amplifier. 
The amplified pulse is compressed with high temporal quality due to the fine dispersion tuning with the CFBG. 
The system can generate a maximum of $\sim$250 nJ of single-pulse energy with an average power of $\sim$1~W at a repetition rate of 4~MHz. The laser system is suitable for multiphoton microscopy.

\end{abstract}

%
\vspace{2pc}
%
%
%
%

\section{Introduction}
Fiber laser systems operating within the short-wavelength infrared region (SWIR). (1.0–2.0 {\textmu}m) have attracted significant interest due to their importance in multi-photon microscopy (MPM) \cite{horton2013vivo, wang2018comparing,nomura2020short,murakoshi2022vivo}, communication systems \cite{sakamoto199635, yamada2013tm3+} and optical spectroscopy \cite{ono2015broadband, wilson2015review}. 
In MPM, the nonlinear excitation process requires high peak intensities, which are achieved using ultrashort laser pulses. To avoid photodamage and heating in biological tissues, the average power must be kept low, which can be managed through appropriate choices of pulse energy and repetition rate.
Pulse durations of $\sim$200~fs are commonly used, as they enhance the probability of simultaneous multiphoton absorption while avoiding the dispersion and handling challenges associated with too short pulses in microscopes \cite{xu2024multiphoton}. 
The compact and simple design of fiber lasers is favored for easier integration in bio-imaging experiments. 
Therefore, developing a fiber laser system that is both simple and capable of delivering sufficient average power with reasonable pulse duration and repetition rate is highly desirable. 
For effective three-photon excitation of most widely used green and red fluorescent proteins, excitation wavelengths within the two spectral windows (at approximately 1.3 {\textmu}m and 1.7 {\textmu}m) are typically used.  
In fluorescence-based MPM, the 1.7 {\textmu}m band is preferred, as it offers a window where the combined effects of absorption and scattering are minimized in most biological tissues compared to the 1.3~{\textmu}m band \cite{guesmi2018dual,takano2025watt}. 

However, a significant challenge in fiber laser development lies in identifying a suitable gain medium capable of efficient operation within this specific SWIR window, while simultaneously achieving direct pulse generation and amplification sufficient for MPM. 
In Ref.~\cite{khegai2018nalm}, high-germania bismuth active fiber was able to operate in 1.7 {\textmu}m. However, the pulse duration is too long ($\ge$ 600 fs) for MPM. Thulium-doped (Tm) fibers are well-capable of the direct generation of ultrashort pulses in the target wavelength as they exhibit a broad emission, spanning 1.6 to 2.1 {\textmu}m owing to the $^3F_4$ to $^3H_6$ energy-level transition of Tm ions \cite{nomura2016efficient, zhang2023recent}. 
However, for MPM applications, some ingenuity is required to mitigate the undesirable amplified spontaneous emission (ASE) at long wavelength region ($\geq$ 1.8 {{\textmu}}m) of Tm emission which can damage samples in \textit{in vivo} imaging applications. 

Several approaches have been reported for suppressing longer wavelengths, although each suffers from inherent limitations. 
For example, one can design a photonic crystal fiber to suppress the propagation of long wavelength components.
Although this approach enables a wide tuning range, the resulting pulse durations remain on the order of picoseconds, rendering it unsuitable for MPM applications \cite{emami20171700}. 
Furthermore, the implementation of photonic crystal structures requires sophisticated fabrication processes.

Chen et al. demonstrated short-pass filtering near 1.7~{\textmu}m using W-type normal-dispersion Tm-doped fibers \cite{chen2021w,chen20211,chen2025high}. In their work, femtosecond pulses were generated through a bend-induced wave-selection technique to suppress longer wavelengths. However, the reported pulses exhibited pronounced pedestals that carried a considerable fraction of the pulse energy. These pedestals are likely caused by strong nonlinear effects accumulated during propagation through excessively long fiber lengths, which distort the phase of ultrashort pulses. Consequently, the use of free-space components remains necessary to mitigate these distortions.




In this paper, we present a multistage ultrafast amplifier system based on Tm:Tb:ZBLAN fibers. Unlike in Ref.\cite{emami20171700} which requires an advanced fabrication process or in 
Ref. \cite{chen2025high} which depends on mechanical bending  for short-pass filtering effect, our technique is based on ionic codoping. Ionic codoping provides an intrinsic and fiber-integrated mechanism for suppressing longer wavelengths. This approach avoids excessive nonlinear phase accumulation and eliminates the need for additional free-space filtering components, offering a compact and effective solution for wavelength control in our ultrafast fiber laser system. We doped Tm ions into the fiber core while doping the cladding with Tb ions to effectively suppress amplification in the long-wavelength range \cite{sakamoto199635,yamada2012broadband,yamada2013tm3+}. The absorption spectrum of the Tb ion is centered at $\sim$1.9 {\textmu}m. Essentially, the Tb ions act as an energy sink for the long-wavelength photons emitted by the Tm ions.  Here, we use ZBLAN glass as the host material because it has almost zero dispersion near 1.7 {\textmu}m. This property can lead to high quality pulse compression. A ZBLAN fiber also has a higher efficiency compared to silica fiber when used as an active fiber because its lower phonon energy reduces non-radiative decay, leading to a longer upper-state lifetime, which is crucial for achieving sufficient gain \cite{nomura2020short, walsh2004comparison, eichhorn2008comparative}.
In addition to this, we have also improved our compression stage by incorporating a chirped fiber Bragg grating (CFBG) on-demand paired with a Treacy grating compressor for controlled dispersion management. With these, we have succeeded in generating watt-level pulses  at 1.75 {\textmu}m and compressed pedestal-free pulses of 222 fs duration, which are suitable for MPM.   

\section{Tm:Tb:ZBLAN fiber laser system}
\begin{figure}
    \centering
\includegraphics[width=1\linewidth]{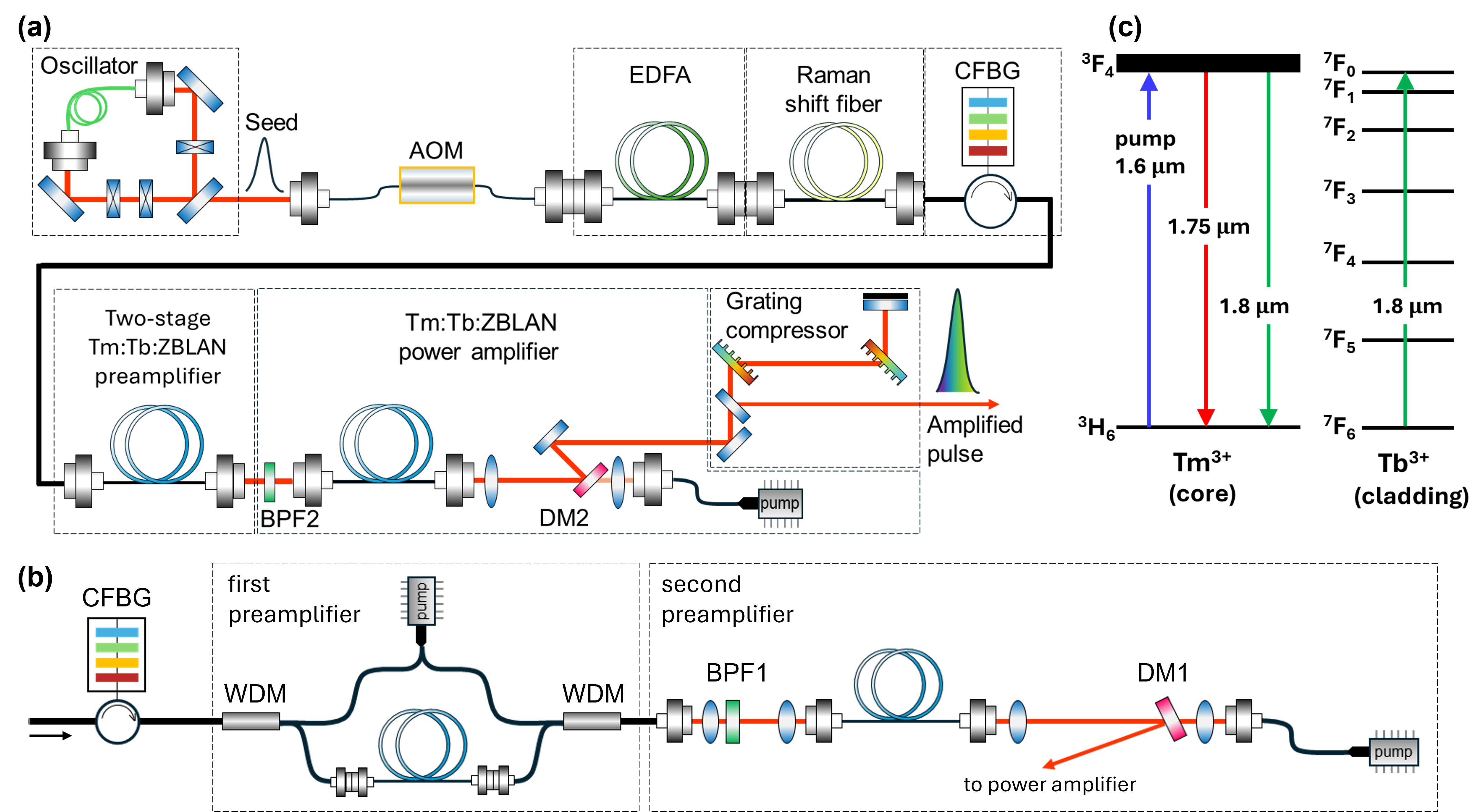}
    \caption{(a) Schematic of the laser system. AOM: acousto-optic modulator, EDFA: Erbium-doped fiber amplifier, CFBG: chirped fiber Bragg grating, BPF: bandpass filter, and DM: dichroic mirror. (b) Schematic of the two-stage preamplifier system. The solid red and black lines indicate free-space and fiber connection beam paths, respectively. (c) Energy level transitions of Tm$^{3+}$ and Tb$^{3+}$.}  
    \label{fig:Fig1}
\end{figure}

\begin{figure}
    \centering
\includegraphics[width=0.8\linewidth]{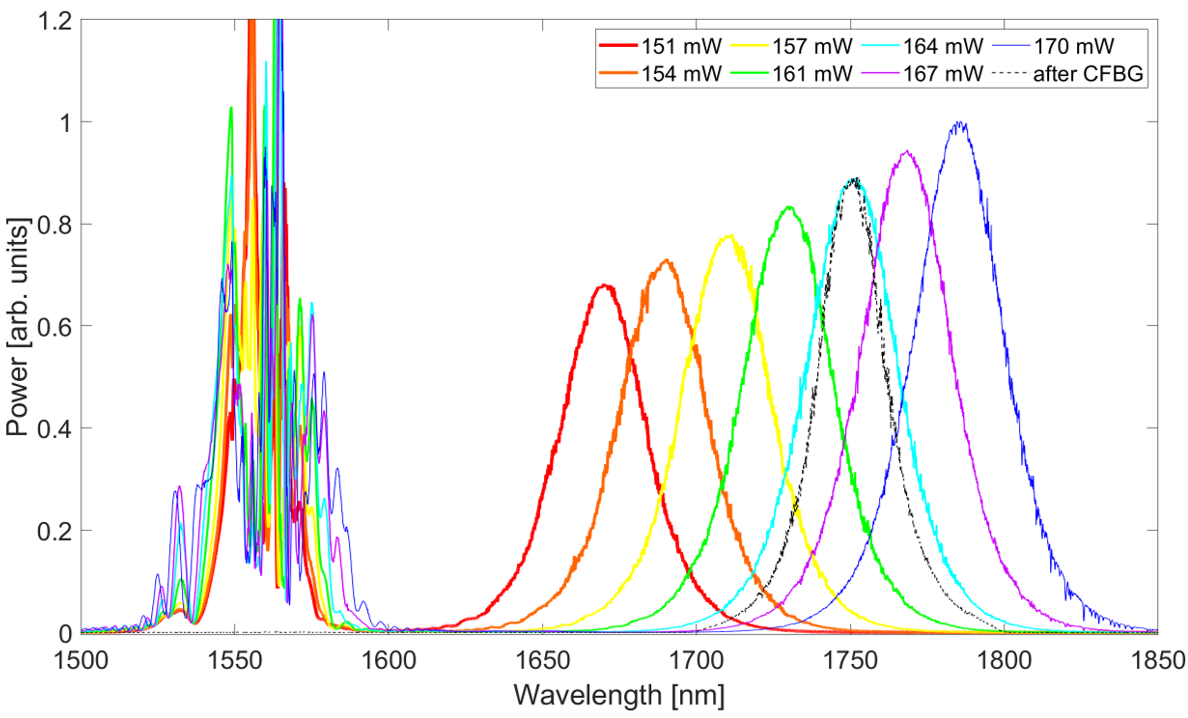}
    \caption{Output spectra of the Raman shift fiber with various pump powers for the EDFA. The spectrum of the pulse reflected by the CFBG is also shown.}
    \label{fig:Fig2}
\end{figure}

Figure~\ref{fig:Fig1}(a) shows the schematic of the laser system based on Tm:Tb:ZBLAN fibers. The system started with a commercially available erbium-doped silica fiber oscillator (VFLP-1560-M-fs-FA, Connet).  
Initially, the pulse repetition frequency, as determined by the cavity length was 50 MHz and was reduced to $\sim$4 MHz using an acousto-optic modulator (AOM). The pulse was amplified with a commercial polarization-maintained erbium-doped fiber amplifier. The 1.75~{\textmu}m seed was generated through soliton self-frequency shift of the amplified 1560 nm pulses. The Raman shift fiber was a small cladding polarization-maintained silica fiber (RCHA15-PS-U17C, MFD:6.0 {\textmu}m @1550 nm, Fujikura) with a length of 5 m. Figure~\ref{fig:Fig2} shows the spectra from the Raman shift fiber with various pump currents. The transform limited duration of the 1.75 {\textmu}m soliton is 95 fs by assuming a sech$^2$ pulse shape. The 1.75 {\textmu}m component was estimated to be 8.7~mW. Then it was sent to a CFBG-based tunable pulse stretcher (\textit{TPSR, TeraXion}) which introduces an initial group delay dispersion (GDD) and third-order dispersion (TOD) of $+$7.1435 ps$^2$ and $-$0.0585 ps$^3$, respectively, and can be tuned. The tuning range of our CFBG is 0.0525 ps$^2$ for GDD and  0.0027 ps$^3$ for TOD.

The CFBG stretches the pulse to $\sim$112 ps. The spectrum after the CFBG is slightly narrower than that of the soliton due to the diffraction bandwidth of the CFBG (see Figure~\ref{fig:Fig2}). The average power of the stretched pulse was estimated to be 0.35~mW. The low output power after the CFBG is due to the insertion losses of the input and output ports of the circulator in addition to the 30$\%$ reflectivity loss of the CFBG. The total loss within the CFBG and the circulator is 6.6 dB. 

The stretched pulse was sent to a two-stage Tm:Tb:ZBLAN fiber preamplifier, whose output is further amplified with a power amplifier. The details of the first and second preamplifiers are shown in Figure~\ref{fig:Fig1}(b). These three stages of Tm:Tb:ZBLAN fiber amplifiers are used to scale up the average power. Essentially, direct amplification of the 1.75 {\textmu}m in our multistage amplifier system relies on the $^3F_4$ to $^3H_6$ energy-level transition of Tm ions. T. Sakamoto et al. first designed and fabricated a Tm–Tb-doped fiber, which had a core doped with Tm and a cladding doped with Tb ions\cite{sakamoto199635}. As illustrated in Figure~\ref{fig:Fig1}(c), the Tb-doped cladding drives the Tm$^{3+}$ gain spectrum to the 1.75 {\textmu}m region because the 1.8–2 {\textmu}m ASE of Tm$^{3+}$ is well suppressed by the ground state absorption of Tb$^{3+}$ \cite{zhang2023recent}. 

\begin{figure}
    \centering
    \includegraphics[width=1\linewidth]{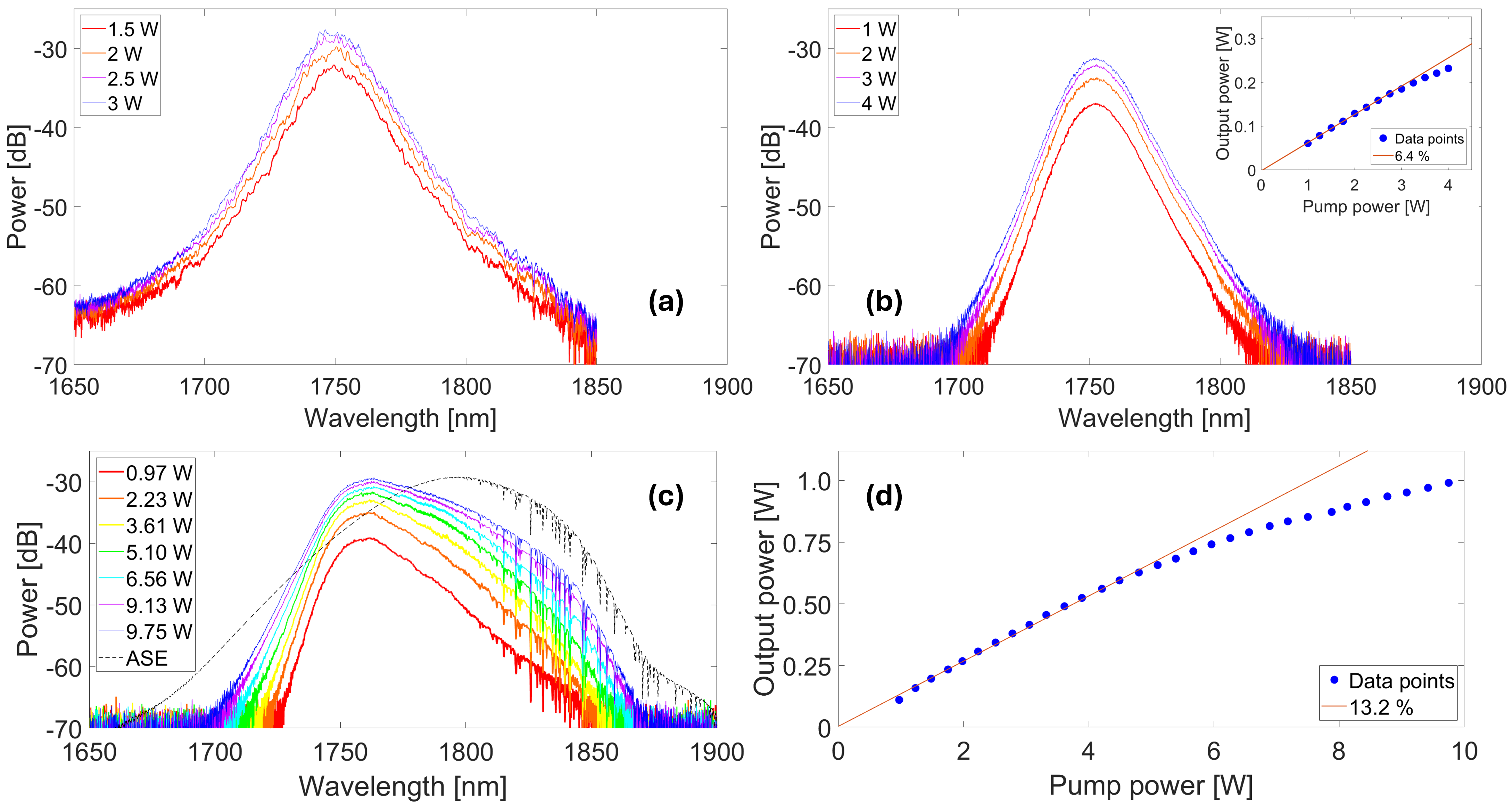}
    \caption{(a) Output spectra of the first Tm:Tb:ZBLAN preamplifier at various pump powers. (b) Output spectra of the second Tm:Tb:ZBLAN preamplifier at several pump powers. (c) Power spectra using the 4m gain fiber and (d) slope efficiency of the final power amplifier at different pump powers. The ASE spectrum of the Tm:Tb:ZBLAN fiber at 9.75 W is also shown in (c) alongside the amplified spectrum.}
    \label{fig:Fig3}
\end{figure}

The first preamplification stage is based on fiber coupling. The Tm:Tb:ZBLAN fiber is pumped by a continuous wave 1600 nm Er:SiO$_{2}$ fiber laser (VFLS-1600-M-5-FA, Connet). The diameter of the core, the numerical aperture (NA), the length, the Tm and Tb doping concentration of the fiber were 8.5 {\textmu}m, 0.14, 1.6 m,  2500 ppm and 3000~ppm, respectively. The ZBLAN fiber was connected to wavelength division multiplexing (WDM) based on SMF28 silica fibers with optical contact. The losses of the input and output ports for the seed pulse were 0.8 dB and 0.9 dB, respectively. The Tb ions in the cladding have zero absorption around 1600 nm \cite{sakamoto199635}. Hence, we do not expect significant loss during amplification due to this design. The spectra of the amplified stimulated emission with several pump powers are shown in Figure~\ref{fig:Fig3}(a). The average power of the amplified pulses was 12 mW with 3 W pumping at a repetition rate of 4 MHz. The slope efficiency was 0.6 $\%$.
The pulse was further amplified using a second stage preamplifier after the pulse passed through a bandpass filter (BPF1) (BP-1693-140). 

The second preamplifier consists of a Tm:Tb:ZBLAN with a core diameter of 7.5~{\textmu}m, NA of 0.14, a doping concentration of 2500 ppm for Tm and 3000 ppm for Tb and a length of 3 m. While the first preamplifier is based on fiber coupling, the second stage is based on space coupling to avoid too much nonlinear effect during propagation of the amplified pulse. The pump laser is the same 1600 nm Er:SiO\textsubscript{2} fiber laser (VFLS-1600-B-5-FA, Connet). 
The output spectra of the second amplifier is shown in Figure~\ref{fig:Fig3}(b) until 4 W of pumping power. We did not observe an obvious change in the shape of the amplified spectrum though it is slightly red-shifted by $\sim 3$ nm compared to the output of the first stage. This may be due to Tm:Tb:ZBLAN having higher gain at longer wavelengths. Nonetheless, it is very unlikely that such a very small shift affects the compression quality of the final output. The central wavelength of the spectrum peaked at 1.75 {\textmu}m which is shorter than in our previous system (1.77 {\textmu}m)~\cite
{okada2024femtosecond}. The output power of the second preamplifier with 4 W pumping was 232 mW, which is almost twice higher than our previous report (130 mW) at a 4 MHz repetition rate~\cite{okada2024femtosecond}. 
The slope efficiency is 6.4 $\%$.  
The second preamplifier was improved due to the use of a longer active fiber, which enhances population inversion and thus a higher gain. The noise in the long wavelength region ($\geq$1.8 {\textmu}m) is also suppressed very well with an almost 30 dB power level difference to the central peak. 

The power amplifier also consisted of a Tm:Tb:ZBLAN fiber with a core diameter of 7.6 {\textmu}m, NA of 0.14, a length of 4 m, and a doping concentration of 2500 ppm for Tm and 3000 ppm for Tb.  The spectrum of the ASE of the active fiber with the pump power of 9.75~W is shown in black dash line in Figure~\ref{fig:Fig3}(c). To ensure that we only amplify the target 1.75 {\textmu}m wavelength and filter out otherwise dominant long wavelength emission ($\geq$1.8 {\textmu}m), we inserted a bandpass filter (BPF2) (BP-1693-140) at the output of the second preamplifier. The BPF takes the function of the $4f$ system in our previous system but now with lesser footprint \cite{okada2024femtosecond}. The pump laser for the power amplifier is a commercially available Raman shift fiber laser (RLR-30-1620, IPG photonics). The output spectrum of the power amplifier is shown in Figure \ref{fig:Fig3}(c) for several pump powers. The ASE was suppressed well as the amplified pulse peaked around 1.76~{\textmu}m. We see some intensity in the long wavelength region ($\geq$ 1.8 {\textmu}m) but we can easily remove it at the compression stage without losing too much power. With respect to the second preamplifier spectrum, the output spectrum is slightly shifted towards the longer wavelength region where the gain is higher. No obvious spectral broadening or narrowing was observed, indicating that nonlinear effects were well controlled during the amplification process. 
The pump power dependence of the output power was linear up to 4.21~W and was slightly saturated. 
The slight saturation observed in the main amplifier at high pump power may be attributed to misalignment during coupling of the pump laser into the fiber. Specifically, at increased pump power, thermal effects could induce a shift in the fiber position, leading to reduced coupling efficiency.
The slope efficiency was calculated to be 13.2 $\%$ by fitting the output power up to the 4.21~W absorbed pump power. At 9.75 W pump power, the output power reached $\sim$1~W. 
We estimated $\sim$250 nJ of single pulse energy. The output power and the slope efficiency can still be improved by boosting the seed pulse. 
As shown in the inset in Figure~\ref{fig:Fig3}(b), the gain of the Tm:Tb:ZBLAN fiber in the second stage has not reached the saturation point, implying that the output power can be further increased with higher pumping power. 

Unlike in Ref. \cite{okada2024femtosecond}, we used Tm:Tb:ZBLAN fiber in the power amplifier similar to the preamplifiers to reduce the complexity of the laser system. To our knowledge, this is the first all-Tm:Tb:ZBLAN fiber system designed to directly amplify watt-level ultrafast pulses at 1.76~{\textmu}m. 
The peak-to-peak fluctuation measurement resulted in an rms value of 6$\%$ at 20000 counts, which was recorded with a high speed InGaAs detector and a 1 GHz digital oscilloscope. 
Since the amplified pulse came out from a  single mode fiber, the amplified beam had an almost Gaussian profile. The polarization of the final output can be controlled by using a half-wave plate and a quarter-wave plate. 

\section{Pulse compression results}
Previously, we used a fiber for pulse stretching \cite{okada2024femtosecond}. This is not ideal because the introduced dispersion is already fixed and cannot be tuned once it is fabricated.  We also utilized a pair of grating-prism before, where it is possible to compensate for the higher order of dispersion \cite{forget2012transmission}. However, the added degrees of freedom, i.e., the distance between the grating and prism, complicates the optimization of the compression system. 
The current trend is the use of a CFBG that introduces a controlled amount of chromatic dispersion to an incident pulse \cite{bartulevicius2017compact}. 
The tunable CFBG allows us to compensate for timing distortions caused by compressor misalignment and to correct nonlinear effects introduced by the amplifier.
Here, we report the use of a CFBG and a grating pair in tandem as well as the dispersion matching between these elements. We use grating pairs only for pulse compression which is the standard for chirped pulse amplification (CPA) systems \cite{treacy1969optical,strickland1985chirped}. The manufactured transmission gratings (\textit{Gitterwerk}) in our system are capable of delivering extremely high diffraction efficiency.

\begin{figure}
    \centering
    \includegraphics[width=1\linewidth]{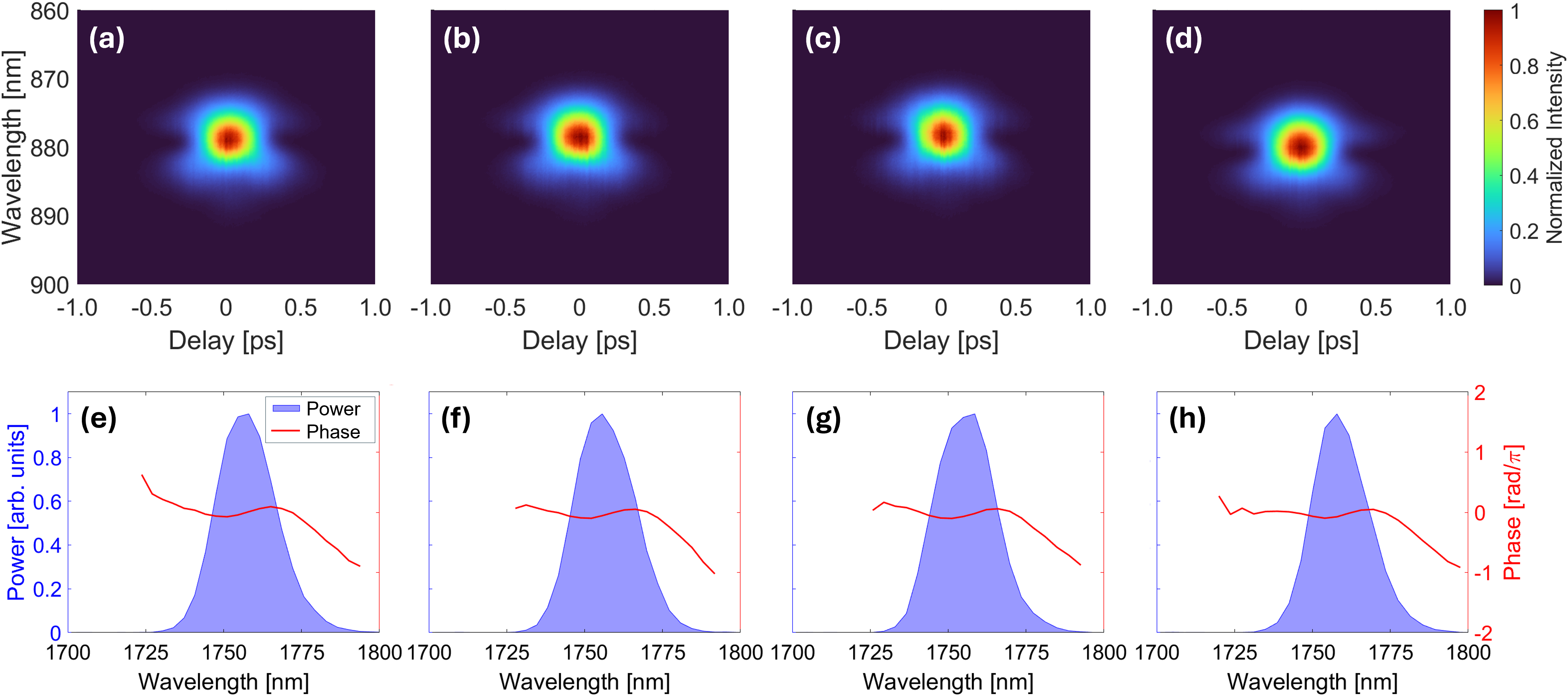}
    \caption{Normalized SHG-FROG traces (top row) of the output pulses from the power amplifier, with their corresponding retrieved spectral profile and phases (bottom row), for pump powers of (a,e) 2.23 W, (b,f) 5.10 W, (c,g) 9.13 W, and (d,h) 9.75 W.}
    \label{fig:Fig4}
\end{figure}

The dispersion of our CFBG is determined by taking into account the complete group delay response of the grating compressor and the total GDD of the amplifier system. The silica fibers of the fiber coupled devices (AOM, WDM, etc. in total 10~m of SMF28) contribute $\sim$ $-$0.5 ps$^2$, whereas the ZBLAN fiber (1.6 m, 3.0 m and 4.0~m of Tm:Tb:ZBLAN) contributes insignificantly to the net dispersion that we do not take into account. 
The theoretically specified distance between the gratings in the Treacy compressor is 214 mm, the groove number of the gratings is 800 mm$^{-1}$ and the incidence angle of the grating (Littrow angle) is 44.4$^\circ$.  The two gratings are arranged in Littrow angle to ensure a best diffraction efficiency of $97\%$ and a roof mirror is used to create a double-pass configuration.  Theoretically, the gratings introduce $-$7.15~ps$^2$ of second-order dispersion for a 1.75 {\textmu}m incident beam. Based on these parameters, the designed dispersion of our CFBG is $+$7.1435 ps$^2$ and $-$0.0585 ps$^3$ for GDD and TOD, respectively. The TPSR implements a CFBG to achieve large dispersive pulse stretching within a compact fiber length. It is spliced with a circulator which allows the pulse to enter the stretcher and then reflect off the CFBG while ensuring that the returning, stretched pulse is rerouted to the output instead of back toward the laser source \cite{tosi2018review}. This enables an all-fiber, single-pass configuration without introducing back-reflections that could destabilize the laser.


%


At first, we investigated the effect of different pumping powers on the pulse duration of the retrieved spectrum for the case when the CFBG is not yet-fine tuned. We only adjusted the grating pair distance L to obtain the shortest pulse duration and to minimize the GDD. The output of the grating pair is characterized using second harmonic generation frequency-resolved optical gating (SHG-FROG, \textit{FemtoEasy}). The results are shown in Figure~\ref{fig:Fig4}.
We found that the optimal distance between the gratings was 174~mm and was constant for all pumping powers. There are no significant differences in the measured pulse duration for different pump powers.  This indicates that nonlinear effects are well-balanced by the dispersion in the system. The durations of the compressed pulses for 2.23 W, 5.10 W, 9.13 W, and 9.75 W were 248 fs, 241 fs, 238 fs, and 243~fs, respectively.  
From each retrieved spectral phase, we calculated the TOD at each pump power and determined that the absolute dispersion between 2.23~W and 9.75~W is sufficiently small ($\sim10^{-5}$~ps\textsuperscript{3}). 
Hence, no evident phase distortion or nonlinear effects were observed in the SHG signal. Specifically, for a pump power of 9.75 W, as shown in Figure~\ref{fig:Fig4}(d,h), we compressed the pulse to 243 fs. Although we have not fine-tuned the CFBG, the pulse duration is already shorter than that of our previous system (254~fs) \cite{okada2024femtosecond}. Moreover, the retrieved spectrum has a very low intensity in the $\geq$ 1.8~{\textmu}m region compared to Ref. \cite{murakoshi2022vivo}. This means that we have mostly suppressed the ASE and that the output power of the compressed pulse came purely from the coherent part of the amplified pulse.  That is also confirmed by a small residual phase within $\pm$ 0.5$\pi$ retrieved from SHG-FROG.



\begin{figure}
    \centering
    \includegraphics[width=1\linewidth]{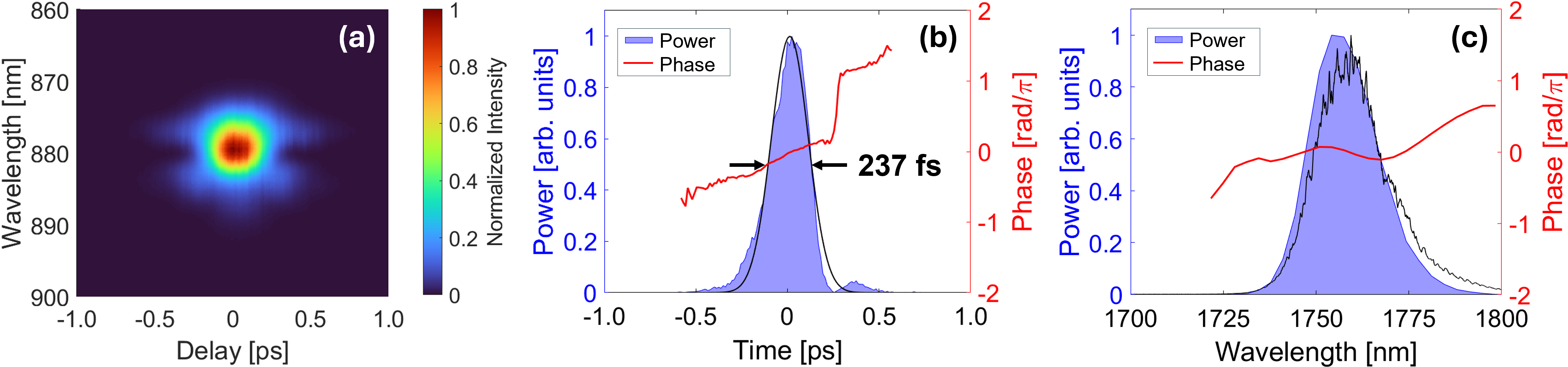}
    \caption{Pulse characterization result at a pump power of 9.75 W with CFBG fine-tuning. (a) The SHG-FROG trace of the compressed pulse by the grating pair using a 4 m Tm:Tb:ZBLAN gain fiber. Retrieved pulse in (b) time and (c) frequency domain. The black solid line in (c) is the spectrum after the compressor. The fine structures on the spectra are due to the multimode fiber used with the spectrometer.}
    \label{fig:Fig5}
\end{figure}

Finally, we examine the effect of the CFBG tuning using a constant pumping power of 9.75 W and a fixed grating distance of 174 mm. We fine-tuned the CFBG by applying offsets in the dispersion slope from $-$90 ps/nm$^2$ to $+$90 ps/nm$^2$. The best FROG trace of the compressed pulse is shown in Figure~\ref{fig:Fig5}(a).
The intensity and phase of the pulse retrieved in the time and frequency domains are also shown in Figures~\ref{fig:Fig5}(b) and ~\ref{fig:Fig5}(c), respectively.  
Applying a dispersion offset of $-$50~ps/nm$^2$, equivalent to a GDD of $+$7.1296~ps$^2$ and a TOD of $-$0.0578~ps$^3$, resulted in a minimum pulse duration of 237~fs. 
The transform-limited pulse duration derived from the spectrum was 210 fs. 
The close match between the transform-limited and retrieved pulse duration proves that the CFBG dispersion profile was successfully compensated to that of the grating pair. 
An excellent temporal quality of the compressed pulse was also achieved, as the retrieved temporal profile shows high contrast.  There is no considerable energy distributed on the sides of the central pulse.  
Therefore, we expect that multi-photon excitation now will be more efficient. 
Although some residual TOD can be seen in the spectral phase, we did not observe any serious phase distortion due to the nonlinear effects in the power amplifier. The retrieved spectral profile of the compressed pulses peaked near 1.76~{\textmu}m, which is blue-shifted relative to our previous system \cite{okada2024femtosecond}. 
It is consistent with the independently measured spectrum immediately after the Treacy compressor (recorded using a AQ6376 $Yokogawa$ spectrometer), as depicted by the black solid plot overlapped in Figure~\ref{fig:Fig5}(c). 
Overall, the consistency between the compressor-output spectrum and the FROG-retrieved spectrum demonstrates that the SHG-FROG accurately captures the dominant spectral content of the pulses. 
The pulse power after the Treacy compressor was estimated to be 700~mW.

\section{Amplification and compression results for the short Tm:Tb:ZBLAN fiber}

We also investigated the amplification and compression characteristics of pulses generated using a shorter gain fiber. In the new power amplifier configuration, the length of the gain medium was reduced to 1.6 m. The ion concentrations were maintained at 2500 ppm and 3000 ppm for the Tm and Tb ions, respectively, and the NA is 0.14. However, the shorter fiber had a larger core diameter of 8.5~{\textmu}m compared to 7.6~{\textmu}m for the 4 m fiber. All other fiber components and laser components remained similar to those with the 4 m fiber.

 \begin{figure}
    \centering
    \includegraphics[width=1\linewidth]{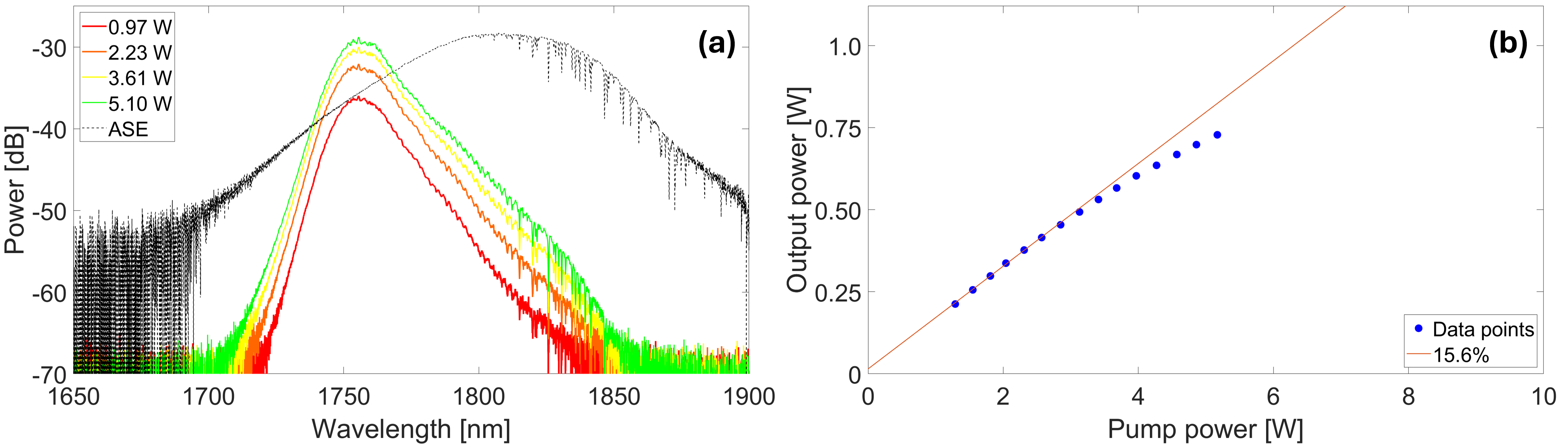}
    \caption{(a) Power spectra using the 1.6 m gain fiber and (d) slope efficiency of the final power amplifier at different pump powers. The ASE spectrum of the Tm:Tb:ZBLAN fiber at 5.10 W is also shown in (a) alongside the amplified spectrum.}
    \label{fig:Fig6}
\end{figure}

 We anticipated that the amplified pulse spectrum would be broader when using the 1.6 m fiber compared to the 4 m fiber, because shorter fiber exhibits a broader ASE spectrum range. However, a comparison of Figure~\ref{fig:Fig3}(c) and Figure~\ref{fig:Fig6}(a) shows otherwise. We note that the apparent broadening in Figure~\ref{fig:Fig3}(c) arises primarily from long-wavelength components and unsuppressed ASE during amplification. In contrast, the short fiber yields a higher signal-to-noise ratio after amplification compared to the long gain fiber. Specifically, for the 4 m fiber, the spectral component at 1.75~{\textmu}m is only about 5 dB higher than that at 1.8~{\textmu}m, as shown in Figure~\ref{fig:Fig3}(c), while for the 1.6 m fiber this difference increases to nearly 20 dB, as depicted in Figure~\ref{fig:Fig6}(a).

Similar to the 4 m fiber case, no obvious spectral narrowing or broadening was observed, indicating that nonlinear effects were negligible. The output power increased linearly with absorbed pump power up to 3.61 W, beyond which slight saturation was observed. By fitting the output power up to 3.61 W, the slope efficiency was calculated to be 15.6$\%$. At an absorbed pump power of 5.10 W, the output power reached 0.7 W and we estimated $\sim$175 nJ of single pulse energy. The slight improvement in efficiency observed with the short fiber can be attributed to its larger core diameter. 

Essentially, in a small-core, long-fiber amplifier (4 m fiber), the pump light is absorbed very strongly near the input end of the fiber. This leads to a high population inversion at the beginning, but the pump is depleted quickly, causing the inversion to drop significantly as the light propagates. 
As a result, a large portion of the fiber toward the output can become weakly inverted or even absorbing for the signal. The extended fiber length also provides a large gain path for ASE, allowing ASE to build up and compete with the signal, which reduces overall efficiency. In contrast, a large-core, short-fiber amplifier (1.6 m fiber) operates with a lower pump intensity per unit area, but the pump absorption is spread more evenly along the fiber length. This leads to a higher average population inversion throughout the active region rather than a strong inversion localized at the input. Because the fiber is shorter, there is less opportunity for ASE growth and signal reabsorption, so more of the stored energy is efficiently transferred to the desired signal. 

\begin{figure}
    \centering
    \includegraphics[width=1\linewidth]{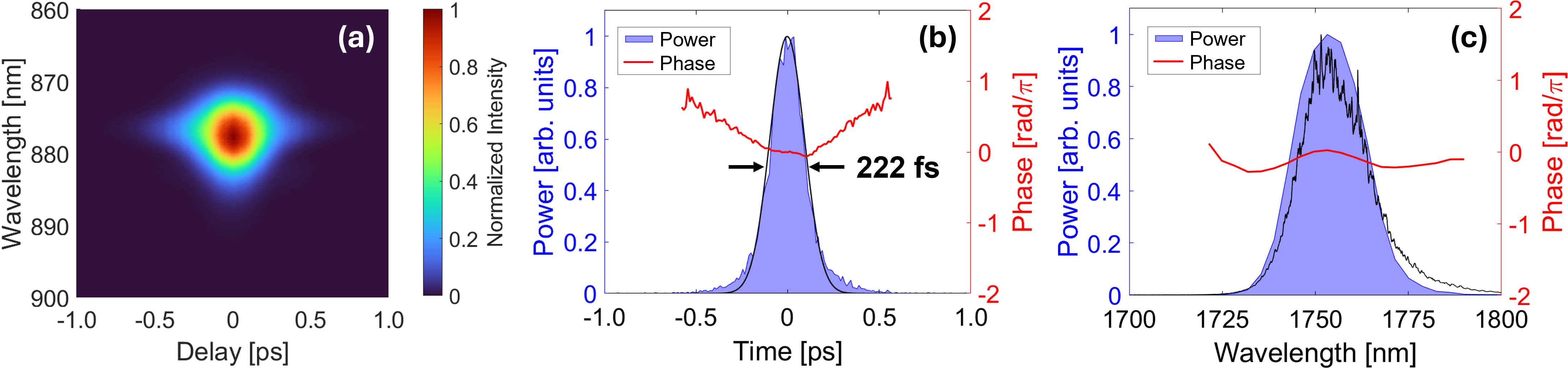}
    \caption{Pulse characterization result at a pump power of 5.10 W with CFBG fine-tuning using a 1.6 m Tm:Tb:ZBLAN gain fiber. (a) The SHG-FROG trace of the compressed pulse by the grating pair. Retrieved pulse in (b) time and (c) frequency domain. The black solid line in (c) is the spectrum after the compressor. The fine structures on the spectra are due to the multimode fiber used with the spectrometer.}
    \label{fig:Fig7}
\end{figure}

Finally, the amplified pulses generated by the 1.6 m fiber were compressed using the same grating-pair compressor. The GDD was reduced by varying the separation of the gratings, and the optimal distance was found to be L = 183 mm. The best FROG trace of the compressed pulse, shown in Figure~\ref{fig:Fig7}(a), was obtained with a dispersion offset corresponding to a GDD of $+$7.1348~ps$^2$ and a TOD of $-$0.0583 ps$^3$. Under these conditions, the shortest measured pulse duration was 222 fs as presented in Figure~\ref{fig:Fig7}(b). 
The transform-limited pulse duration derived from the spectrum was 212 fs. The close agreement between the transform-limited and retrieved pulse durations indicates that the CFBG dispersion profile was effectively compensated for by the grating pair. As depicted in Figure~\ref{fig:Fig7}(c), we were able to
compress the pulse with reasonable quality, as confirmed by a small residual spectral phase. 
The retrieved compressed pulses exhibit a spectral peak near 1.75~\textmu m, in good agreement with the independently measured spectrum recorded immediately after the Treacy compressor, as shown by the solid black line spectrum overlapped in Figure~\ref{fig:Fig7}(c). The spectra recorded with the $Yokogawa$ spectrometer (right after the grating compressor) and FROG are almost the same for the 4~m and 1.6~m fibers and there is no obvious difference in their retrieved spectral phases, as depicted in Figure~\ref{fig:Fig5}(c) and Figure~\ref{fig:Fig7}(c), respectively. 

It should be noted that although the 4 m fiber produced a broader spectrum after the power amplifier with a spectral width of 38 nm, the corresponding compressed pulse duration was not significantly different from that obtained using the 1.6 m fiber with a spectral width of 19 nm. 
This is because the long-wavelength components and incoherent ASE present in the 4 m fiber output do not contribute effectively to pulse compression. Moreover, for the 4 m fiber, the spectra measured before and after the gratings showed a noticeable difference. Specifically, the long-wavelength components of the spectrum, or the ASE, appear attenuated after the grating compressor compared to the spectrum before the compressor. This decrease may be attributed to the random polarization of the ASE and the polarization-dependent nature of our transmission gratings (i.e., s-polarized). Essentially, when the gain medium is pumped to achieve population inversion, spontaneous emission occurs in different directions with varying polarizations. Because our gratings preferentially transmit s-polarized light, some wavelengths are diffracted efficiently for one polarization, while others are poorly diffracted. As a result, the ASE appears to be attenuated after the grating compressor.




\section{Summary and conclusions}
In summary, we have improved our previous CPA system by redesigning the amplification stage and incorporating a CFBG for tunable dispersion control in the compression stage. With the controllable CFBG, the pulse was compressed with reasonable spectral quality, as the spectrogram showed independence from the pumping power, and with an improved temporal profile, as it is free from side artifacts and strong nonlinear effects.
We have also highlighted the characteristics of pulses generated using a long fiber with a small core (4~m fiber) and a short fiber with a large core (1.6~m fiber). For both cases, we achieved watt-level average power and transform-limited duration of $\sim$210~fs. 
We found that the 4 m fiber produced a broader amplified spectrum, which was largely associated with long-wavelength ASE that did not contribute to effective pulse compression. In contrast, the 1.6 m fiber provided a higher signal-to-noise ratio, better suppression of ASE, and cleaner amplified spectrum. Although, in general, the 1.6 m fiber delivers improved pulse quality, we think that these comparisons between the 4 m and 1.6 m gain fibers add caution when designing fiber-based amplifier laser system. We also believe this is the first all-Tm:Tb:ZBLAN amplifier system that directly generates watt-level, short-wavelength ultrafast pulses. The construction of the all-Tm:Tb:ZBLAN amplifier system offers a simpler design while still delivering high average power and pulse duration near 200 fs, making it a suitable light source for three-photon lifetime imaging microscopy experiments with red fluorescent proteins \cite{cheng2014measurements,liu2018ex,prevedel2025three}.





\section*{Funding} National Institute for Physiological Sciences (24NIPS225, 25NIPS233); National Institute of Natural Sciences (OML0123008,OML012410) 

\section*{References}
\bibliography{iopart-num}

\end{document}